\documentclass[12pt]{article}
\title{On representations and differences of Stieltjes coefficients, and other relations} 
\author{Mark W. Coffey\\
Department of Physics\\
Colorado School of Mines\\
Golden, CO  80401\\
(Received $\mbox{~~~~~~~~~~~~~~~~~~~~~~~~~~~~~~~2008}$)}
\date{December 8, 2008}
\begin{document}
\maketitle
\baselineskip=25 pt
\begin{abstract}

The Stieltjes coefficients $\gamma_k(a)$ arise in the expansion of the Hurwitz zeta function $\zeta(s,a)$ about its single simple pole at $s=1$ and are of 
fundamental and long-standing importance in analytic number theory and other
disciplines.  We present an array of exact results for the Stieltjes coefficients,
including series representations and summatory relations.   Other integral 
representations provide the difference of Stieltjes coefficients at rational
arguments.  The presentation serves to link a variety of topics in analysis
and special function and special number theory, including logarithmic series,
integrals, and the derivatives of the Hurwitz zeta and Dirichlet $L$-functions at 
special points.  The results have a wide range of application, both theoretical
and computational.

\end{abstract}
 
\vspace{.25cm}
\baselineskip=15pt
\centerline{\bf Key words and phrases}
\medskip 

\noindent

Stieltjes constants, Dirichlet $L$-series, Hurwitz zeta function, Riemann zeta function, Laurent expansion, functional equation, Gamma function, digamma function, polygamma
function, logarithmic series, harmonic numbers 

\vfill
\centerline{\bf MSC numbers}
11M35, 11M06, 11Y60 
 
\baselineskip=25pt
\pagebreak
\medskip
\centerline{\bf Introduction}
\medskip

The Stieltjes (or generalized Euler) constants $\gamma_k(a)$ appear as 
expansion coefficients in the Laurent series about $s=1$ for the Hurwitz zeta function 
$\zeta(s,a)$, one of the generalizations of the Riemann zeta function $\zeta(s)$.
Elsewhere \cite{coffeyprsa}, we developed new summatory relations amongst
the values $\gamma_k(a)$ as well as demonstrated one of the very recent conjectures put 
forward by Kreminski \cite{kreminski} on the relationship between $\gamma_k(a)$ and
$-\gamma_k(a+1/2)$ as $k\to \infty$ \cite{coffeystieltjes}. 


In this paper, we present an array of exact results for the Stieltjes constants.
These include individual and summatory relations for paired differences of these
coefficients for rational arguments.  Our work provides a unification of several important topics of analysis and analytic number theory.  These include certain logarithmic sums,
integrals of analytic number theory, special functions, the derivatives of the Hurwitz
zeta and Dirichlet $L$-functions at special points, and differences of the Stieltjes constants.  As the corresponding logarithmic sums are slowly converging, our results provide useful complements and alternatives to numerical computation.  In addition, our analytic results and accompanying representations of $\gamma_k(a)$ provide a basis for inequalities and monotonicity results for the Stieltjes constants.

We may stress the fundamental and long-standing nature of the Stieltjes coefficients.  
They arise in the expansion of the Hurwitz zeta function about its {\em unique} polar
singularity.  They can be used to write other important constants of analytic number
theory, and they appear often in describing error terms and as a result of applying
asymptotic analyses.  
As well, they can be expected to play a role in investigations of the
nonvanishing of $L$-functions and their derivatives along the line $s=1$ and 
elsewhere in the critical strip.

An expository paper of Vardi \cite{vardi} discusses the evaluation of certain
logarithmic integrals (or their equivalents through change of variable) and describes
the underlying connection with Dirichlet $L$-series.  However, the presentation is
illustrative and no connection with logarithmic series or the Stieltjes coefficients
is mentioned.  Much more recently, Medina and Moll \cite{medina} have followed Vardi's approach and given a number of examples for integrands containing a rational function.
Additionally, Adamchik \cite{adamchik} considered differences of the first derivative of the Hurwitz zeta function at rational arguments and related them to logarithmic
integrals.  Below, we explicitly write one of his results in terms of a difference of
values of $\gamma_1(a)$.

After exhibiting some basic relations for $\zeta(s,a)$ and $\gamma_k(a)$, we
provide both general relations to logarithmic sums and instances of logarithmic
integrals in terms of the Hurwitz zeta function and iterated logarithmic integrals
in terms of differences of Stieltjes constants.  Motivated in part by such
relations, we then proceed to other new analytic results, systematized as
Propositions.  We provide selected examples for differences of low-order Stieltjes coefficients.  However, our methods apply to $\gamma_k(a)$ for any order $k$ 
for rational arguments $a$.   We further obtain series representations of the
Stieltjes coefficients.  Among these results, we indicate how families of 
rapidly convergent summation representations may be obtained, that are applicable
to computation.

\pagebreak
\centerline{\bf Preliminary definitions and relations}

The Hurwitz zeta function, initially defined by
$$\zeta(s,a)=\sum_{n=0}^\infty {1 \over {(n+a)^s}}, ~~~~\mbox{Re} ~s > 1,
\eqno(0.1)$$
can be analytically continued to the whole complex plane $C-\{1\}$,
satisfying, for all $s$, the functional equation \cite{hansen,ivic}
$$\zeta(1-s,p/q)={{2\Gamma(s)} \over {(2\pi q)^s}} \sum_{r=1}^q \cos \pi
\left ({s \over 2} - {{2rp} \over q}\right ) \zeta(s,r/q),  \eqno(0.2)$$
where $p$ and $q$ are integers and $\Gamma$ is the Gamma function.
Proof of the Hurwitz formula \cite{titch,watson}
$${{(2\pi)^s} \over {2\Gamma(s)}}\zeta(1-s,a) = \cos{{\pi s} \over 2}
\sum_{n=1}^\infty {{\cos 2\pi na} \over n^s} + \sin{{\pi s} \over 2} 
\sum_{n=1}^\infty {{\sin 2\pi na} \over n^s}, ~~~~\mbox{Re} ~s > 1, 
\eqno(0.3)$$
has been simplified in References \cite{berndt} and \cite{yue}.

The defining relation for the Stieltjes constants in terms of a Laurent
expansion is
$$\zeta(s,a)={1 \over {s-1}}+\sum_{k=0}^\infty {{(-1)^k \gamma_k(a)} \over
k!} (s-1)^k, ~~~~~~ s \neq 1. \eqno(0.4)$$
This equation reflects that $\zeta(s,a)$ has a simple pole at $s=1$ with residue $1$.
By convention, $\gamma_k$ represents $\gamma_k(1)$ and thus explicitly we
have \cite{briggs,hardy,kluyver,mitrovic,stieltjes,wilton}
$$\zeta(s)={1 \over {s-1}}+\sum_{k=0}^\infty {{(-1)^k \gamma_k} \over
k!} (s-1)^k, ~~~~~~ s \neq 1. \eqno(0.5)$$
As noted in \cite{wilton} and again in \cite{yue}, we have $\gamma_0(a)=
-\psi(a)$, where $\psi=\Gamma'/\Gamma$ is the digamma function, so that
$\gamma_0=\gamma$, the Euler constant.
A representation for the Stieltjes constants is given by \cite{berndt}
$$\gamma_k(a)=\lim_{N \to \infty} \left [\sum_{j=0}^N {{\ln^k(j+a)} \over  
{j+a}} - {{\ln^{k+1}(N+a)} \over {(k+1)}} \right ].  \eqno(0.6)$$
As written, Eq. (0.6) is not well suited for numerical purposes.


A very recently derived representation \cite{coffeyoctst} (Proposition 3a) that is 
useful for computation and other purposes is given by
$$\gamma_k(a)={1 \over {2a}}\ln^k a-{{\ln^{k+1} a} \over {k+1}} + {2 \over a}
\mbox{Re} \int_0^\infty {{(y/a-i)\ln^k (a-iy)} \over {(1+y^2/a^2)(e^{2 \pi y}
-1)}}dy, ~~~~~\mbox{Re}~ a>0.  \eqno(0.7)$$

In the following, we let $s(n,m)$ denote the Stirling numbers of the first
kind and $C_k(a) \equiv \gamma_k(a) -(\ln^k a)/a$.  With $B_n(x)$ the
Bernoulli polynomials, their periodic extension is denoted 
$P_n(x) \equiv B_n(x-[x])$.  Two other integral representations of the Stieltjes
constants of interest in this paper are \cite{yue} 
$$C_n(a) = (-1)^{n-1} n!\sum_{k=0}^{n+1} {{s(n+1,n+1-k)} \over {k!}}
\int_1^\infty P_n(x-a) {{\ln^k x} \over x^{n+1}} dx, ~~n \geq 1. 
\eqno(0.8)$$
and 
$$\gamma_n(a)=\sum_{k=0}^m {{\ln^n (k+a)} \over {k+a}}-{{\ln^{n+1} (m+a)}
\over {n+1}}-{{\ln^n(m+a)} \over {2(m+a)}}+\int_m^\infty P_1(x)f_n'(x)dx,$$
$$~~~~~~~~~~~~~~~~~~~~~~~~~~~~0<a\leq 1, ~~~~m, n=0,1,2, \ldots ,  \eqno(0.9)$$
where $f_n(x) \equiv \ln^n(x+a)/(x+a)$.

From Eq. (0.4) we have
$$(-1)^\ell[\gamma_\ell(a)-\gamma_\ell(b)]=\left({\partial \over {\partial s}}
\right)^\ell[\zeta(s,a)-\zeta(s,b)]|_{s=1}.  \eqno(0.10)$$
Therefore, from Eq. (0.1) we obtain the relation to logarithmic sums
$$\gamma_\ell(a)-\gamma_\ell(b)=\sum_{n=0}^\infty \left[{{\ln^\ell(n+a)} \over 
{n+a}}-{{\ln^\ell(n+b)} \over {n+b}}\right].  \eqno(0.11)$$
We may stress here a theme of this paper, namely that in Eqs. (0.10) and (0.11)
polar singularities have been precisely eliminated.

Various integrals may written in terms of differences of Hurwitz zeta functions.
For example, we have
$$\int_0^1 {{t^{a-1}\ln^{s-1} t} \over {1+t}}dt=\left(-{1 \over 2}\right)^s \Gamma(s)
\left[\zeta\left(s,{{a+1} \over 2}\right)-\zeta\left(s,{a \over 2}\right)\right],
~~~~\mbox{Re}~s>0, ~~\mbox{Re}~a>0, \eqno(0.12)$$
and 
$$\int_0^1 {{t^{a-1}\ln^{s-1} t} \over {1+t^p}}dt=\left(-{1 \over {2p}}\right)^s \Gamma(s)
\left[\zeta\left(s,{{a+p} \over {2p}}\right)-\zeta\left(s,{a \over {2p}}\right)\right],
~~~~\mbox{Re}~s>0, ~~\mbox{Re}~a>0, ~~\mbox{Re}~p >0.  \eqno(0.13)$$
Equation (0.12) is easily proved by writing the integrand factor $(1+t)^{-1}$ as
geometric series, interchanging summation and integration, and integrating termwise.
The interchange is readily justified on the basis of uniform convergence.  Equation 
(0.13) may be similarly obtained, or through the use of the change of variable
$u=t^p$ and the application of Eq. (0.12).

Differentiation of Eq. (0.12) or (0.13) with respect to the parameters $a$, $p$, and/or
$s$ gives families of integrals.  Specifically, differentiating Eq. (0.13) $j$ times
with respect to $s$ gives
$$\int_0^1 {{t^{a-1}\ln^{s-1} t} \over {1+t^p}}\ln^j \ln t ~dt
=\left({\partial \over {\partial s}}\right)^j \left(-{1 \over {2p}}\right)^s \Gamma(s)
\left[\zeta\left(s,{{a+p} \over {2p}}\right)-\zeta\left(s,{a \over {2p}}\right)\right],$$
$$~~~~~~~~~~~~~~~~~~~~~~~~\mbox{Re}~s>0, ~~\mbox{Re}~a>0, ~~\mbox{Re}~p >0.  \eqno(0.14)$$
In this way, we may relate such integrals to sums of differences of Stieltjes constants.
As an example special case we have
$$\int_0^1 {{t^{a-1}\ln(\ln t)} \over {1+t^p}}dt=\left. {\partial \over {\partial s}} \right|_{s=1} \left(-{1 \over {2p}}\right)^s \Gamma(s)
\left[\zeta\left(s,{{a+p} \over {2p}}\right)-\zeta\left(s,{a \over {2p}}\right)\right]$$
$$={1 \over {2p}}[\gamma-i\pi+\ln(2p)]\left[\psi\left({a \over {2p}}\right)-
\psi\left({{a+p} \over {2p}}\right)\right]-{1 \over {2p}}\left[\gamma_1\left({a \over
{2p}}\right)-\gamma_1\left({{a+p} \over {2p}}\right)\right].  \eqno(0.15)$$

A result of Adamchik \cite{adamchik} (Proposition 1) holds for integers $p$ and $q$
with $p/q<1$.  It states that
$$\zeta'\left(1,1-{p \over q}\right)-\zeta'\left(1,{p \over q}\right)=
\gamma_1\left(1,{p \over q}\right)-\gamma_1\left(1,1-{p \over q}\right)$$
$$=\pi \cot\left( {{p \pi} \over q}\right)[\ln(2\pi q)+\gamma]-2\pi \sum_{j=1}^{q-1}
\ln \Gamma\left({j \over q}\right)\sin\left({{2\pi j p} \over q}\right).  \eqno(0.16)$$
This result was obtained on the basis of differentiating Rademacher's formula and
is attributed to Almkvist and Meurman.

\medskip
\centerline{\bf Summatory relation for the difference of Stieltjes constants}
\medskip

We have
{\newline \bf Proposition 1}.  We have the summatory relation
$$\sum_{n=0}^\infty {1 \over {n!}}[\gamma_{n+1}(a)-\gamma_{n+1}(b)]=\ln\left[
{{\Gamma(b)} \over {\Gamma(a)}}\right], ~~~~\mbox{Re} ~a>0, ~~\mbox{Re} ~b>0.  
\eqno(1.1)$$

Proof.  We supply two different proofs.  For the first, we apply the relation
\cite{sri} (pp. 87, 92)
$$\zeta'(0,a)=\ln \Gamma(a)- \ln \sqrt{2\pi}.  \eqno(1.2)$$
From Eq. (0.4) we have
$$\zeta'(s,a)=-{1 \over {(s-1)^2}}+\sum_{n=0}^\infty {{(-1)^{n+1}} \over {n!}} 
\gamma_{n+1}(a)(s-1)^n.  \eqno(1.3)$$
Then putting $s=0$ and using Eq. (1.2) we have
$$\zeta'(0,a)-\zeta'(0,b)=\ln\left[{{\Gamma(a)} \over {\Gamma(b)}}\right], \eqno(1.4)$$
and the Proposition follows.

For the second proof we initially assume that $a>0$ and $b>0$.  
We apply the representation (0.7) for $\gamma_k(a)$ and have
$$\sum_{k=0}^\infty {1 \over {k!}}\gamma_{k+1}(a)={1 \over 2}\ln a+a-a\ln a-1
+{2 \over a}\mbox{Re}\int_0^\infty {{(y/a-i)(a-iy)\ln(a-iy)} \over {(1+y^2/a^2)
(e^{2\pi y}-1)}}dy$$
$$=a+\left({1 \over 2}-1-a\right)\ln a+2 ~\mbox{Im} \int_0^\infty {{\ln(a-iy)} \over
{(e^{2\pi y}-1)}}dy, \eqno(1.5)$$
where we have used elementary exponential sums, and Im $\ln(a-iy)=-\tan^{-1}(y/a)$.  
As $\int_a^b \psi(s)ds=\ln[\Gamma(b)/\Gamma(a)]$, the Proposition follows from
Binet's second expression for $\ln \Gamma(z)$ \cite{sri} (pp. 17, 91), 
$$\ln \Gamma(z)=\left(z-{1 \over 2}\right)\ln z - z +{1 \over 2}\ln (2\pi)
+2 \int_0^\infty {{\tan^{-1}(t/z)} \over {e^{2\pi t}-1}}dt, ~~~~\mbox{Re} ~z>0.
\eqno(1.6)$$
By analytic continuation, the result (1.1) is extended to Re $a>0$ and Re $b>0$.

{\bf Remarks}.  The representation (0.9) or its equivalents may also be used to prove Proposition 1, in view of \cite{edwards} (p. 107)
$$\ln \Gamma(s+1)=(s+1/2)\ln s - s+{1 \over 2}\ln 2\pi-\int_0^\infty {{P_1(x)} \over {(x+s)}} dx, ~~~~~\mbox{Re} ~s>0. \eqno(1.7)$$

The representation (0.7) may be used to develop many other summations, including for
$|z|\leq 1$,
$$\sum_{k=0}^\infty {z^k \over {k!}}\gamma_{k+1}(a)={1 \over 2}a^{z-1} \ln a+(a^z-1)
{1 \over z^2}-{a^z \over z}\ln a+2 ~\mbox{Im} \int_0^\infty {{(a-iy)^{z-1}} \over
{(e^{2\pi y}-1)}}\ln(a-iy)dy. \eqno(1.8)$$

Since $\zeta'(-1)=1/12-\ln A$, where is $A$ is Glaisher's constant, we find from Eq.
(1.3) that 
$$\sum_{n=0}^\infty {2^n \over {n!}}\gamma_{n+1}=\ln A- {1 \over 3}<0.  \eqno(1.9)$$

The polygamma functions $\psi^{(j)}$ are connected to the Hurwitz zeta function via
$\psi^{(n)}(x)=(-1)^{n+1}n! \zeta(n+1,x)$ for integers $n \geq 1$.  Therefore, we 
obtain from Eq. (0.4) for the trigamma function
$$\psi'(a)=1+\sum_{k=0}^\infty {{(-1)^k} \over {k!}}\gamma_k(a), \eqno(1.10)$$
and more generally
$$\psi^{(n)}(a)=(-1)^{n+1}n!\left[{1 \over n}+\sum_{k=0}^\infty {{(-1)^k} \over {k!}}\gamma_k(a)n^k\right], ~~~~n \geq 1.  \eqno(1.11)$$
This equation may be taken as an infinite linear system.  Its inversion would yield
the Stieltjes coefficients in terms of polygammic constants.

\medskip
\centerline{\bf Relation to derivatives of Dirichlet $L$-functions}
\medskip

We now introduce Dirichlet $L$-functions $L_{\pm k}(s)$ (e.g., \cite{ireland}, Ch. 16), that are known to be expressible as linear combinations of Hurwitz zeta functions.  
We let $\chi_k$ be a real Dirichlet character modulo $k$, where the corresponding $L$ 
function is written with subscript $\pm k$ according to $\chi_k(k-1)=\pm 1$.  We have
$$L_{\pm k}(s)=\sum_{n=1}^\infty {{\chi_k(n)} \over n^s}={1 \over k^s}\sum_{m=1}^k 
\chi_k(m) \zeta\left(s,{m \over k}\right), ~~~~\mbox{Re} ~s >1.  \eqno(2.1)$$
This equation holds for at least Re $s>1$.  If $\chi_k$ is a nonprincipal character, 
as we typically assume in the following, then convergence obtains for Re $s>0$.

The $L$ functions, extendable to the whole complex plane, satisfy the functional
equations \cite{zucker76}
$$L_{-k}(s)={1 \over \pi}(2\pi)^s k^{-s+1/2}\cos\left({{s \pi} \over 2}\right)\Gamma
(1-s)L_{-k}(1-s), \eqno(2.2)$$
and
$$L_{+k}(s)={1 \over \pi}(2\pi)^s k^{-s+1/2}\sin\left({{s \pi} \over 2}\right)\Gamma
(1-s)L_{+k}(1-s). \eqno(2.3)$$
Owing to the relation
$$\Gamma(1-s)\Gamma(s)={\pi \over {\sin \pi s}}, \eqno(2.4)$$
these functional equations may also be written in the form
$$L_{-k}(1-s)=2(2\pi)^{-s} k^{s-1/2}\sin\left({{\pi s} \over 2}\right) \Gamma(s)
L_{-k}(s), \eqno(2.5)$$
and
$$L_{+k}(1-s)=2(2\pi)^{-s} k^{s-1/2}\cos\left({{\pi s} \over 2}\right) \Gamma(s)
L_{+k}(s). \eqno(2.6)$$

Integral representations are known for these $L$-functions.  In particular, we have
{\newline \bf Lemma 1} \cite{zucker76}.  For $\chi_k$ a nonprincipal Dirichlet 
character we have
$$L_{\pm k}(s)={1 \over {\Gamma(s)}}\int_0^\infty {u^{s-1} \over
{1-e^{-ku}}} \left(\sum_{m=1}^k \chi_k(m)e^{-mu}\right) du, ~~~~\mbox{Re} ~s>0.  \eqno(2.7)$$

{\bf Remark}.  We may note that convergence obtains in the stated domain due to
contribution in the numerator arising from cancellation due to the nonprincipal
character.

Proof.  For Re $s>1$ and Re $a>0$, we have the integral representation (e.g.,
\cite{sri}, p. 89)
$$\zeta(s,a)={1 \over {\Gamma(s)}}\int_0^\infty {{t^{s-1} e^{-(a-1)t}} \over
{e^t-1}} dt.  \eqno(2.8)$$
We use Eq. (2.1) to write
$$L_{\pm k}(s)={{k^{-s}} \over {\Gamma(s)}}\sum_{m=1}^k \chi_k(m) \int_0^\infty
{{t^{s-1} e^{-(m/k-1)t}} \over {e^t-1}}dt.  \eqno(2.9)$$
We then sum the finite geometric series and put $u=t/k$, giving the Lemma for
Re $s>1$.  By analytic continuation for $\chi_k$ nonprincipal, it also holds for
Re $s>0$.

Key relations upon which we build are contained within the following.
{\newline \bf Proposition 2}.  For $\chi_k$ a nonprincipal Dirichlet character 
we have (a)
$$\left.{\partial \over {\partial s}}L_{\pm k}(s)\right|_{s=1}=
-\sum_{n=2}^\infty {{\chi_k(n)\ln n} \over n} \eqno(2.10)$$
$$=k^{-1}\sum_{m=1}^k \chi_k(m)\left[\zeta'\left(1,{m \over k}\right)-\ln k ~\zeta
\left(1,{m \over k}\right)\right] \eqno(2.11)$$
$$=k^{-1}\sum_{m=1}^k \chi_k(m)\left[\gamma_1\left({m \over k}\right)+\ln k ~\psi
\left({m \over k}\right)\right] \eqno(2.12)$$
$$=\int_0^\infty {{(\ln u+\gamma)} \over {1-e^{-ku}}}\left(\sum_{m=1}^k \chi_k(m)
e^{-mu}\right) du, \eqno(2.13)$$
and (b)
$$\left.{\partial^2 \over {\partial s^2}}L_{\pm k}(s)\right|_{s=1}=
\sum_{n=2}^\infty {{\chi_k(n)\ln^2 n} \over n} \eqno(2.14)$$
$$=k^{-1}\sum_{m=1}^k \chi_k(m)\left[\zeta''\left(1,{m \over k}\right)-2\ln k ~\zeta'
\left(1,{m \over k}\right)+\ln^2 k ~\zeta\left(1,{m \over k}\right)\right] \eqno(2.15)$$
$$=k^{-1}\sum_{m=1}^k \chi_k(m)\left[\gamma_2\left(1,{m \over k}\right)-2\ln k ~\gamma_1
\left(1,{m \over k}\right)-\ln^2 k ~\psi\left(1,{m \over k}\right)\right] \eqno(2.16)$$
$$=\int_0^\infty {{[\ln^2 u+2\gamma \ln u+\gamma^2-\zeta(2)]} \over {1-e^{-ku}}}\left(\sum_{m=1}^k \chi_k(m) e^{-mu}\right) du. \eqno(2.17)$$

Proof.  For part (a), Eqs. (2.10) and (2.11) follow from Eq. (2.1).  For Eq. (2.12),
we differentiate Eq. (2.1) and use the important relation for $\chi_k$ nonprincipal
$$\sum_{r=1}^{k-1}\chi_k(r)=\sum_{r=1}^k \chi_k(r)=0.  \eqno(2.18)$$
Evaluation of the result at $s=1$ gives Eq. (2.12).  For Eq. (2.13) we have from
Lemma 1
$${\partial \over {\partial s}}L_{\pm k}(s)= {1 \over {\Gamma(s)}}
\int_0^\infty {{u^{s-1}[\ln u-\psi(s)]} \over {1-e^{-ku}}}\left(\sum_{m=1}^k \chi_k(m)
e^{-mu}\right) du, \eqno(2.19)$$
where we used $\Gamma'(s)=\Gamma(s)\psi(s)$.  Evaluating at $s=1$, with $\psi(1)=-
\gamma$, gives Eq. (2.13).

Part (b) follows very similar steps.  In obtaining Eq. (2.17), we use
$\psi'(1)=\zeta(2)=\pi^2/6$, where $\psi'$ is the trigamma function.

{\bf Remark}.  A key feature of Proposition 2 connecting the values $L_{\pm k}'(1)$
with differences of Stieltjes constants is the nullification of polar singularities.
An example of various relations of this Proposition, Eq. (0.11) at $\ell=1$, and
Eq. (0.15) at $p=1$ is given by
$$\gamma_1\left({a \over 2}\right)-\gamma_1\left({{a+1} \over 2}\right)=\left.
{\partial \over {\partial s}}\right|_{s=1} 2^s \sum_{n=0}^\infty {{(-1)^n} \over
{(n+a)^s}} \eqno(2.20a)$$
$$=\ln 2\left[\psi\left({{a+1} \over 2}\right)-\psi\left({a \over 2}\right)\right]
+2\sum_{n=0}^\infty (-1)^{n+1}{{\ln(n+a)} \over {(n+a)}}.  \eqno(2.20b)$$
When $a=1/2$, the Dirichlet $L$-function appearing on the right side of Eq. (2.20a)
is $L_{-4}(s)$.  

\medskip
\centerline{\bf Case of $\chi_k(k-1)=-1$}
\medskip


We have
\newline {\bf Proposition 3}.  Suppose that $\chi_k$ is a nonprincipal character 
and that $\chi_k(k-1)=-1$.  Then
$$\sum_{m=1}^k \chi_k(m) \gamma_1\left({m \over k}\right)=kL_{-k}'(1)-
\ln k ~ \sum_{m=1}^k \chi_k(m)\psi \left({m \over k}\right)$$
$$=k\int_0^\infty {{(\ln u+\gamma)} \over {1-e^{-ku}}}\left(\sum_{m=1}^k \chi_k(m)
e^{-mu}\right) du-\ln k ~ \sum_{m=1}^k \chi_k(m)\psi \left({m \over k}\right)$$
$$=-{\pi \over k^{1/2}}(\ln 2\pi +\gamma)\sum_{m=1}^k m \chi_k(m)-\pi k^{1/2} \ln
\prod_{m=1}^k \Gamma^{\chi_k(m)} \left({m \over k}\right)
-\ln k ~ \sum_{m=1}^k \chi_k(m)\psi \left({m \over k}\right).  \eqno(2.21)$$

Proof.  The first two equalities in Proposition 3 follow directly from Proposition 2.
For the third, we express $L_{-k}'(1)$ in terms of known quantities of $L_{-k}'(0)$
from the functional equation (2.2) or (2.5).  Given that $\zeta(0,a)=1/2-a$, we have
from Eq. (2.1) that
$$L_{\pm k}(0)=\sum_{m=1}^k \chi_k(m)\left({1 \over 2}-{m \over k}\right)
=-{1 \over k}\sum_{m=1}^k m \chi_k(m), \eqno(2.22)$$
where we used property (2.18).  Parallel to Eq. (2.11) we obtain
$$\left.{\partial \over {\partial s}}L_{\pm k}(s)\right|_{s=0}
=k^{-1}\sum_{m=1}^k \chi_k(m)\left[\zeta'\left(0,{m \over k}\right)-\ln k ~\zeta
\left(0,{m \over k}\right)\right]$$
$$=-\ln k ~L_{\pm k}(0)+\sum_{m=1}^k \chi_k(m) \ln \Gamma\left({m \over k}\right),
\eqno(2.23)$$
where we used both Eqs. (1.2) and (2.18).  From the functional equation (2.5) we find
$$-L'_{-k}(0)={k^{1/2} \over \pi}(\ln k-\ln 2\pi-\gamma)L_{-k}(1)+{k^{1/2} \over \pi}
L_{-k}'(1).  \eqno(2.24)$$
This equation can be solved for $L_{-k}'(1)$, where Eq. (2.23) gives $L_{-k}'(0)$
and by the functional equation (2.2) we have $L_{-k}(1)=(\pi/k^{1/2})L_{-k}(0)$.
The result is
$$L_{-k}'(1)=-{\pi \over k^{3/2}}(\ln 2\pi+\gamma)\sum_{m=1}^k m\chi_k(m)-{\pi \over
k^{1/2}} \ln \prod_{m=1}^k \Gamma^{\chi_k(m)} \left({m \over k}\right), \eqno(2.25)$$
yielding, by the first line of Eq. (2.21), the conclusion of the Proposition.

We note that the digamma function, like the Gamma function itself, satisfies a 
number of identities that may be used to re-express Eq. (2.21).  These include the
reflection formula
$$\psi(z)-\psi(1-z)=-\pi \cot \pi z, \eqno(2.26)$$
as well as the multiplication formula for integers $m$,
$$\psi(mz)=\ln m+\sum_{k=0}^{m-1} \psi\left(z+{k\over m}\right).  \eqno(2.27)$$
In addition, we recall that by means of Gauss's formula (\cite{sri}, p. 19) the
value $\psi(p/q)$ for any rational argument 
can be written as a finite combination of elementary function values.  As concerns
differences of higher order Stieltjes coefficients, similar identities may be
written for the trigamma and higher order polygamma functions.

\medskip
{\bf Examples}
\medskip

As examples of Proposition 3, we may write the following, using Eq. (2.26).

$$\gamma_1\left({1\over 3}\right)-\gamma_1\left({2\over 3}\right)=-{\pi\over \sqrt{3}}\left\{\ln 2\pi+\gamma-3\ln\left[{{\Gamma\left({1 \over 3}\right)} \over
{\Gamma\left({2 \over 3}\right)}}\right]+\ln 3\right\}, \eqno(2.28)$$

$$\gamma_1\left({1\over 4}\right)-\gamma_1\left({3\over 4}\right)=-\pi\left\{\ln 8\pi+\gamma-2\ln\left[{{\Gamma\left({1 \over 4}\right)} \over
{\Gamma\left({3 \over 4}\right)}}\right]\right\}, \eqno(2.29)$$ 

$$\gamma_1\left({1\over 6}\right)-\gamma_1\left({5\over 6}\right)=\pi\left\{2\sqrt{
{2 \over 3}}(\ln 2\pi+\gamma)-\sqrt{6}\ln\left[{{\Gamma\left({1 \over 6}\right)} \over
{\Gamma\left({5 \over 6}\right)}}\right]+\sqrt{6}\ln 6 \right\}, \eqno(2.30)$$ 

$$\gamma_1\left({1\over {7}}\right)+\gamma_1\left({2\over {7}}\right)
-\gamma_1\left({3\over {7}}\right)+\gamma_1\left({4\over {7}}\right)
-\gamma_1\left({5\over {7}}\right)-\gamma_1\left({6\over {7}}\right)$$
$$=\sqrt{7}\pi\left\{(\ln 2\pi+\gamma)-\ln\left[{{\Gamma\left
({1 \over {7}}\right)\Gamma\left({2 \over {7}}\right)\Gamma\left
({4 \over {7}}\right)} \over {\Gamma\left({3 \over {7}}\right)\Gamma\left
({5 \over {7}}\right)\Gamma\left({6 \over {7}}\right)}}\right]
+\ln 7\right\}, \eqno(2.31)$$
and
$$\gamma_1\left({1\over {11}}\right)-\gamma_1\left({2\over {11}}\right)
+\gamma_1\left({3\over {11}}\right)+\gamma_1\left({4\over {11}}\right)
+\gamma_1\left({5\over {11}}\right)-\gamma_1\left({6\over {11}}\right)
-\gamma_1\left({7\over {11}}\right)-\gamma_1\left({8\over {11}}\right)
+\gamma_1\left({9\over {11}}\right)-\gamma_1\left({{10} \over {11}}\right)$$
$$=\pi\left\{{1 \over \sqrt{11}}(\ln 2\pi+\gamma)-\sqrt{11}\ln\left[{{\Gamma\left
({1 \over {11}}\right)\Gamma\left({3 \over {11}}\right)\Gamma\left
({4 \over {11}}\right)\Gamma\left({5 \over {11}}\right)\Gamma\left
({9 \over {11}}\right)} 
\over {\Gamma\left({2 \over {11}}\right)\Gamma\left
({6 \over {11}}\right)\Gamma\left({7 \over {11}}\right)\Gamma\left
({8 \over {11}}\right)\Gamma\left({{10} \over {11}}\right)}}\right]
-\ln 11\right\}. \eqno(2.32)$$

In such equations, we could just as well use the duplication formula
$${{\Gamma(x)} \over {\Gamma(2x)}} = \sqrt{\pi}{2^{1-2x} \over {\Gamma(x+1/2)}}
\eqno(2.33)$$
to re-express the Gamma function ratios.  More generally, we may use the
multiplication formula
$$\Gamma(nx)=(2\pi)^{(1-n)/2}n^{nx-1/2}\prod_{k=0}^{n-1} \Gamma\left(x+{k \over n}
\right).  \eqno(2.34)$$  
Proposition 3 shows that the logarithmic sums (0.11) for rational values of $a$ and
$b$ are essentially logarithmic constants.  The integrals corresponding to the examples (2.28)-(2.32) are easily written from Eq. (2.21) and we omit the details.
Comparison can be made to tabulated integrals that are expressible in terms of 
logarithmic ratios of Gamma function values \cite{adamchik,grad,medina,vardi}.  
These include entries on pages 532, 571-573, and 580-581 of a standard table \cite{grad}.

Regarding examples (2.28), (2.31), and (2.32) we may recall that for $k$ an odd prime, there is exactly one nonprincipal Dirichlet character $\chi_k$ modulo $k$.  

By \cite{coffeystieltjes} (Proposition 3) or as a special case of Proposition 5.1 
of \cite{coffeyprsa}, for integers $q \geq 2$ we have
$$\sum_{r=1}^{q-1} \gamma_k\left({r \over q}\right)=-\gamma_k + q(-1)^k {{\ln^{k+1} q} \over {(k+1)}} + q\sum_{j=0}^k {k \choose j} (-1)^j (\ln^j q) \gamma_{k-j}.  
\eqno(2.35)$$
Given Proposition 3, we may now combine various sums and differences
of Stieltjes coefficients to find identities in terms of the values $\gamma_k\equiv
\gamma_k(1)$.  As a very particular instance, we have
{\newline \bf Corollary 1}.  The values $\gamma_1(1/3)$ and $\gamma_1(2/3)$ may be
separately written in terms of $\gamma_1$.

\noindent
This statement follows from Proposition 3 at $k=3$ [Example (2.28)] together with
Eq. (2.35) at $q=3$ and $k=1$.

In the case of $\gamma_1(a)$, the relation (0.16) of Adamchik may also be invoked.

\medskip
\centerline{\bf Case of $\chi_k(k-1)=+1$}
\medskip

We have
\newline {\bf Proposition 4}.  Suppose that $\chi_k$ is a nonprincipal character 
and that $\chi_k(k-1)=+1$.  Then
$$\sum_{m=1}^k \chi_k(m) \gamma_1\left({m \over k}\right)=kL_{+k}'(1)-
\ln k ~ \sum_{m=1}^k \chi_k(m)\psi \left({m \over k}\right)$$
$$=k\int_0^\infty {{(\ln u+\gamma)} \over {1-e^{-ku}}}\left(\sum_{m=1}^k \chi_k(m)
e^{-mu}\right) du-\ln k ~ \sum_{m=1}^k \chi_k(m)\psi \left({m \over k}\right)$$
$$=k^{1/2}\left[2(\ln 2\pi +\gamma)\ln \prod_{m=1}^k \Gamma^{\chi_k(m)} \left({m \over k}
\right)-\sum_{m=1}^k \chi_k(m)\zeta''\left(0,{m \over k}\right)\right]
-\ln k ~ \sum_{m=1}^k \chi_k(m)\psi \left({m \over k}\right).  \eqno(2.36)$$

Proof.  When $\chi_k(k-1)=1$, $L_{+k}(0)=0$, $\sum_{m=1}^k m\chi_k(m)=0$, and 
$$L_{+k}'(0)=\sum_{m=1}^k \chi_k(m)\ln \Gamma\left({m \over k}\right).  \eqno(2.37)$$
Differentiating Eq. (2.1) we have
$$L_{+k}(0)=\sum_{m=1}^k \chi_k(m)\left[-2\ln k ~\ln \Gamma\left({m \over k}\right)
+\zeta''\left(0,{m \over k}\right)\right].  \eqno(2.38)$$
Differentiating the functional equation (2.3) we find
$$L_{+k}'(1)={1 \over k^{1/2}}\left[2\left(\gamma+\ln\left({{2\pi} \over k}\right)\right)
L_{+k}'(0)-L_{+k}''(0)\right].  \eqno(2.39)$$
Then using Eqs. (2.35) and (2.36) we determine
$$L_{+k}'(1)=k^{-1/2}\left[2(\ln 2\pi +\gamma)\ln \prod_{m=1}^k \Gamma^{\chi_k(m)} 
\left({m \over k}\right)-\sum_{m=1}^k \chi_k(m)\zeta''\left(0,{m \over k}\right)\right].
\eqno(2.40)$$
Substituting into the first line of Eq. (2.36), the Proposition is completed.

{\bf Examples}.  We have from Proposition 4 at $k=5$ using Eq. (2.4),
$$\gamma_1\left({1\over {5}}\right)-\gamma_1\left({2\over {5}}\right)
-\gamma_1\left({3\over {5}}\right)+\gamma_1\left({4\over {5}}\right)$$
$$=\sqrt{5}\left\{2(\gamma+\ln 2\pi)\ln {1\over 2}(1+\sqrt{5})-\zeta''\left(0,{1 \over 5}
\right)+\zeta''\left(0,{2 \over 5}\right)+\zeta''\left(0,{3 \over 5}\right) -\zeta''\left(0,{4 \over 5}\right)\right.$$
$$\left. + 2\ln 5 \coth^{-1} \sqrt{5}\right\}.  \eqno(2.41)$$

We have from Proposition 4 at $k=10$ using Eq. (2.4),
$$\gamma_1\left({1\over {10}}\right)-\gamma_1\left({3\over {10}}\right)
-\gamma_1\left({7\over {10}}\right)+\gamma_1\left({9\over {10}}\right)$$
$$=\sqrt{10}\left\{2(\gamma+\ln 2\pi)\ln {1\over 2}(3+\sqrt{5})-\zeta''\left(0,{1 \over {10}}\right)+\zeta''\left(0,{3 \over {10}}\right)+\zeta''\left(0,{7 \over {10}}\right) -\zeta''\left(0,{9 \over {10}}\right)\right.$$
$$\left. + 3\sqrt{2}\ln 10 \coth^{-1} \sqrt{5}\right\}.  \eqno(2.42)$$

From the well known Hermite formula for $\zeta(s,a)$ (e.g., \cite{sri}, p. 91)
$$\zeta(s,a)={a^{-s} \over 2} + {a^{1-s} \over {s-1}} + 2 \int_0^\infty 
{{\sin(s \tan^{-1} y/a)} \over {(y^2+a^2)^{s/2}}} {{dy} \over {(e^{2\pi y}-1)}},  \eqno(2.43)$$
we obtain
$$\zeta''(0,a)={1\over 2}\ln^2a+2a\ln a-a\ln^2a-2a-2\int_0^\infty \tan^{-1}\left({y
\over a}\right) {{\ln(a^2+y^2)} \over {(e^{2 \pi y}-1)}}dy.  \eqno(2.44)$$
Evaluation of the integral for rational values of $a$, or combinations of rational
values of $a$, would be of interest in regard to Proposition 4.  It appears that a
contour integral evaluation may be possible, with the integrand having simple poles
along the imaginary axis at $y=ji$ and residues $(i/2\pi)\tanh^{-1}(j/a)\ln(a^2-j^2)$
there.  
However, this is likely to give an infinite series representation of $\zeta''(0,a)$,
whereas another closed form is desirable.

\medskip
\centerline{\bf An explicit summation expression for the Stieltjes constants}
\medskip

In this section we develop an explicit formula for $\gamma_n(a)$, based
upon the representation (0.9).  What is required is an evaluation of the
logarithmic integral there.   We have
{\newline \bf Proposition 5}.  Let $\Gamma(s,t)$ be the incomplete Gamma function.
Then for $0<a\leq 1$, $m, n=0,1,2, \ldots$, we have
$$\gamma_n(a)=\sum_{k=0}^m {{\ln^n (k+a)} \over {k+a}}-{{\ln^{n+1} (m+a)}
\over {n+1}}-{{\ln^n(m+a)} \over {2(m+a)}}$$
$$+\sum_{j=m}^\infty\left\{\ln^n(j+a+1)-\ln^n(j+a)-{1 \over {(n+1)}}\left[\ln^{n+1}(j+a+1) -\ln^{n+1}(j+a)\right]\right.$$
$$\left.-(a+j+1/2)\left[\Gamma[n,\ln(j+a)]-\Gamma[n,\ln(j+a+1)] -\Gamma[n+1,\ln(j+a)]+\Gamma[n+1,\ln(j+a+1)]\right]\right\}.  \eqno(3.1)$$
Proof.  We may proceed as follows:
$$\int_m^\infty P_1(x)f_n'(x)dx=\sum_{j=m}^\infty \int_j^{j+1}P_1(x)f_n'(x)
dx=\sum_{j=m}^\infty \int_j^{j+1} (x-j-1/2)f_n'(x)dx.  \eqno(3.2)$$
By using the expression for $f_n$, we obtain
$$\int_m^\infty P_1(x)f_n'(x)dx=\sum_{j=m}^\infty \int_{j+a}^{j+a+1}
\left[{1 \over x}-(a+j+1/2){1 \over x^2} \right]\left(n\ln^{n-1} x 
-\ln^n x\right) ~dx$$
$$=\sum_{j=m}^\infty \left\{-(a+j+1/2)\int_{j+a}^{j+a+1}
{{[\ln^{n-1}x-\ln^n x]} \over x^2} dx -{1 \over
{(n+1)}}\left[\ln^{n+1}(j+a+1)-\ln^{n+1}(j+a)\right] \right.$$ 
$$\left. +\ln^n(j+a+1)-\ln^n(j+a) \right\}.  \eqno(3.3)$$
The remaining integrals may be performed as \cite{grad}
$$\int_{j+a}^{j+a+1} {{\ln^n x} \over x^2} dx=\Gamma[n+1,\ln(j+a)]
-\Gamma[n+1,\ln(j+a+1)]. \eqno(3.4)$$
The insertion of Eqs. (3.3) and (3.4) into Eq. (0.9) provides the Proposition.

{\bf Remarks}.  Semi-infinite integrals over $P_1$ as we have just performed are
of much interest in connection with applications of Euler-Maclaurin summation.

In conjunction with Eq. (3.1) we may note that for $n$ a nonnegative integer we have
\cite{grad} (p. 941)
$$\Gamma(n+1,x)=n!e^{-x}\sum_{m=0}^n {x^m \over {m!}}.  \eqno(3.5)$$

We may obtain other explicit summation representations of the Stieltjes
constants by working with Eq.\ (0.8).  Here we show the approach for the
special case of $\gamma_1$.  However, by using the defining relation
$B_n(x)=\sum_{k=0}^n {n \choose k} B_k x^{n-k}$, where $B_j$ are
Bernoulli numbers, thereby introducing another sum, this can be carried out 
generally.

As an example, we have
$$C_1(1)=\gamma_1=-\sum_{k=0}^2 {{s(2,2-k)} \over {k!}}\int_1^\infty
P_1(x) {{\ln^k x} \over x^2} dx,  \eqno(3.6)$$
where $s(2,2)=1$, $s(2,1)=-1$, and $s(2,0)/2=0$.  
The integral of Eq. (3.5) is given by
$$\int_1^\infty P_1(x) {{\ln^k x} \over x^2}dx=\sum_{j=1}^\infty \int_j^{j+1}
\left [{1 \over x}-(j+1/2){1 \over x^2}\right] \ln^k x dx$$
$$=\sum_{j=1}^\infty\left\{ {1 \over {k+1}}\left[\ln^{k+1}(j+1)-\ln^{k+1}
j\right]-(j+1/2)\int_j^{j+1}{{\ln^k x} \over x^2} dx \right \}. \eqno(3.7)$$
Denoting the integral
$I_j(k)=\int_j^{j+1} (\ln^k x)/x^2 dx$, we find by integration by parts
the recursion relation
$$I_j(k)=kI_j(k-1)+(j+1/2)\left[{{\ln^k(j+1)} \over {j+1}}- {{\ln^k j} \over 
j} \right ].  \eqno(3.8)$$
The solution of the homogeneous part of this recursion relation is of course
just $k!I_j(0)$.  Another way to evaluate the integrals $I_j(k)$ is in
terms of the incomplete Gamma function, as in Eq. (3.4).  As Eq. (3.8)
indicates, when effectively $a \to 0$ in Eq. (3.4), $I_j(k)$ can be
expressed explicitly for any fixed power $k$.
The combination of Eqs. (3.6), (3.7), and (3.8) gives an explicit summation
representation of the constant $\gamma_1 \simeq -0.0728158454836767$.

\medskip
\centerline{\bf New expressions for $\gamma_1(a)$ and $\gamma_k(a)$}
\medskip

We have
{\newline \bf Proposition 6}.  Let $H_n=\sum_{k=1}^n 1/k$ be the $n$th harmonic
number.  Then we have for Re $a>0$
$$\gamma_1(a)=-{1 \over 2}\ln^2 (a+1)+\sum_{k=1}^\infty {{(-1)^k} \over {(k+1)}}
[(\zeta(k+1,a)-a^{-(k+1)})H_k +\zeta'(k+1,a)]+\ln a \ln\left(1+{1 \over a}\right).
\eqno(4.1)$$

Proof.  By Eq. (0.6) at $k=1$ we have
$$\gamma_1(a)=\lim_{N \to \infty} \left [\sum_{j=0}^N {{\ln(j+a)} \over  
{j+a}} - {{\ln^2(N+a)} \over 2} \right ]$$ 
$$=\lim_{N \to \infty}\left[\sum_{j=0}^N {{\ln(j+a)} \over  
{j+a}} - \int_{1-a}^N {{\ln(x+a)} \over {x+a}}dx \right ]$$ 
$$=\sum_{j=0}^\infty {{\ln(j+a)} \over {j+a}} - 
\int_{1-a}^1 {{\ln(x+a)} \over {x+a}}dx -\int_1^\infty {{\ln(x+a)} \over {x+a}}dx$$
$$=-{1 \over 2}\ln^2(a+1)+{{\ln a} \over a}+\sum_{j=1}^\infty 
{{\ln(j+a)} \over {j+a}} -\sum_{j=1}^\infty \int_j^{j+1} {{\ln(x+a)} \over {x+a}}dx$$
$$=-{1 \over 2}\ln^2(a+1)+{{\ln a} \over a}+\sum_{j=1}^\infty \int_0^1 \left[
{{\ln(j+a)} \over {j+a}} - {{\ln(x+j+a)} \over {x+j+a}}\right]dx.  \eqno(4.2)$$
We now apply the generating function for harmonic numbers
$$\sum_{n=1}^\infty (-1)^n H_n z^n =-{{\ln(1+z)} \over {1+z}}, ~~~~~~~~|z|< 1, \eqno(4.3)$$
to expand the integrand in this equation.  We have
$${{\ln y} \over y}-{{\ln(x+y)} \over {x+y}}=-{{\ln(1+x/y)} \over {x+y}}-\ln y\left(
{1 \over {x+y}} -{1 \over y}\right)$$
$$=-{{\ln(1+x/y)} \over {x+y}}-{{\ln y} \over y}\left[{1 \over {(1+x/y)}}-1\right]$$
$$=-{1 \over y}{{\ln(1+x/y)} \over {1+x/y}}-{{\ln y} \over y}\sum_{k=1}^\infty (-1)^k
{x^k \over y^k}$$
$$=\sum_{k=1}^\infty (-1)^k {{(H_k-\ln y)} \over y^{k+1}}x^k.  \eqno(4.4)$$

We obtain for the integral in Eq. (4.2)
$$\int_0^1 \sum_{k=1}^\infty (-1)^k {{[H_k-\ln(j+a)]} \over {(j+a)^{k+1}}} x^k dx
=\sum_{k=1}^\infty {{(-1)^k} \over {(k+1)}}{{[H_k-\ln(j+a)]} \over {(j+a)^{k+1}}}.  \eqno(4.5)$$
Summing this expression over $j=1$ to $\infty$ gives 
$$\sum_{k=1}^\infty \sum_{j=1}^\infty {{(-1)^k} \over {(k+1)}}{{[H_k-\ln(j+a)]} \over {(j+a)^{k+1}}}=\sum_{k=1}^\infty {{(-1)^k} \over {(k+1)}}[\zeta(k+1,a+1)H_k+
\zeta'(k+1,a+1)]$$
$$=\sum_{k=1}^\infty {{(-1)^k} \over {(k+1)}}[(\zeta(k+1,a)-a^{-(k+1)})H_k+
\zeta'(k+1,a)+a^{-(k+1)}\ln a], \eqno(4.6)$$
where we used the relation $\zeta(s,a+1)=\zeta(s,a)-a^{-s}$ and its derivative.
Inserting Eq. (4.6) into Eq. (4.2) gives the Proposition.

We mention a second method for proving Proposition 6, by using the integral
representation (2.8) for $\zeta(s,a)$.  This method can also provide an integral
representation of $\gamma_1(a)$.  In particular, since
$$\zeta'(k+1,a)={1 \over {k!}}\int_0^\infty {{t^k e^{-(a-1)t}} \over
{e^t-1}} \ln t ~dt,  \eqno(4.7)$$
we have the term of Eq. (4.1)
$$\sum_{k=1}^\infty {{(-1)^k} \over {(k+1)}} \zeta'(k+1,a)=\int_0^\infty 
{{e^{-(a-1)t}} \over {e^t-1}}\left[{1 \over t}-1-{e^{-t} \over t}\right] \ln t ~dt.  \eqno(4.8)$$
This integral may be evaluated per logarithmic differentiation of 
$$\int_0^\infty t^\beta {{e^{-(a-1)t}} \over {e^t-1}}\left[{1 \over t}-1-{e^{-t} \over t}\right] dt=\Gamma(\beta)[a^\beta -\beta \zeta(\beta+1,a)].  \eqno(4.9)$$
Then one may apply the operator $(\partial/\partial \beta)_{\beta=0}$, use the
functional equation of the Hurwitz zeta function, and the relations $\zeta(0,a)=1/2-a$
and (1.4).

For the other summation term in Eq. (4.1), using (2.8), we have
$$\sum_{k=1}^\infty {{(-1)^k} \over {(k+1)}}\zeta(k+1,a) H_k
=-\int_0^\infty {e^{-(a+1)t} \over {(1-e^{-t})t}}\left[\gamma(1+e^t)+2e^{t/2}\cosh\left(
{t \over 2}\right)\ln t\right.$$
$$\left. -e^{t/2}\sqrt{\pi t}\left.{{\partial I_\nu} \over {\partial \nu}}
\right|_{\nu=-1/2}\left({t \over 2}\right)\right], \eqno(4.10)$$
where $I_\nu$ is the modified Bessel function of the first kind (e.g., \cite{grad},
pp. 958, 961).

{\bf Note 1}.  By integrating the generating function relation (4.3) we obtain the
term of Proposition 6
$$\sum_{k=1}^\infty {{(-1)^k} \over {(k+1)}}a^{-(k+1)}H_k =-{1 \over 2}\ln^2\left(
{{a+1} \over a}\right).  \eqno(4.11)$$
{\bf Corollary 2}.  We have
$$\gamma_1=-{1 \over 2}\ln^2 2+\sum_{k=1}^\infty {{(-1)^k} \over {(k+1)}}
[(\zeta(k+1)-1)H_k +\zeta'(k+1)]$$
$$=\sum_{k=1}^\infty {{(-1)^k} \over {(k+1)}}[\zeta(k+1)H_k +\zeta'(k+1)].  \eqno(4.12)$$
{\newline \bf Corollary 3}.  Proposition 6 permits the recovery of relation (2.35)
at $k=1$,
$$\sum_{r=1}^{q-1} \gamma_1\left({r \over q}\right)=(q-1)\gamma_1-q\left({1 \over 2}+
\gamma\right)\ln q.  \eqno(4.13)$$

\noindent
In verifying this statement we use 
$$\sum_{r=1}^{q-1} \zeta\left(k+1,{r \over q}\right)=(q^{k+1}-1)\zeta(k+1).  \eqno(4.14)$$


Rapidly converging expressions for $\gamma_k(a)$ for a complex domain of $a$ may be obtained by using very recently constructed series representations of the Hurwitz zeta function \cite{coffeyhurwitz}.  As an instance of this we find
{\newline \bf Proposition 7}.  Let the generalized harmonic number 
$H_n^{(2)}=\sum_{k=1}^n 1/k^2=\psi'(1)-\psi'(n+1)=\pi^2/6-\psi'(n+1)$.  Then we have
for Re $a>1/2$
$$\gamma_1(a)=-{1 \over 2}\ln^2\left(a-{1 \over 2}\right)+\sum_{k=1}^\infty {1
\over {4^k(2k+1)}}\left[H_{2k}\zeta(2k+1,a)+\zeta'(2k+1,a)\right], \eqno(4.15)$$
and
$$-\gamma_2(a)={1 \over 3}\ln^3\left(a-{1 \over 2}\right)+\sum_{k=1}^\infty {1 \over {4^k(2k+1)}}\left[(H_{2k}^2-H_{2k}^{(2)})\zeta(2k+1,a)\right.$$
$$\left. +2H_{2k}\zeta'(2k+1,a)+\zeta''(2k+1,a)\right]. \eqno(4.16)$$
More generally, for $k \geq 1$ and $n \geq 1$ let
$$r_n(k,m) \equiv (-1)^{m-n} {{s(2k+1,n-m+1)} \over {(n-m+1)_{m+2k-n}}}, ~~~~
0 \leq m \leq n,  \eqno(4.17)$$
where $(z)_a=\Gamma(z+a)/\Gamma(z)$ is the Pochhammer symbol and $s(n,m)$ are the
Stirling numbers of the first kind.  Then we have
$$\gamma_n(a)=-{1 \over {n+1}}\ln^{n+1}\left(a-{1 \over 2}\right)-(-1)^n\sum_{k=1}^\infty
{1 \over {4^k(2k+1)}}\sum_{m=0}^n {n \choose m}r_n(k,m)\zeta^{(m)}(2k+1,a).  \eqno(4.18)$$

Proof.  We use \cite{coffeyhurwitz} (Corollary 3) for Re $a>1/2$,
$$\zeta'(s,a)=-{{(a-1/2)^{1-s}} \over {s-1}} \ln(a-1/2)-{{(a-1/2)^{1-s}} \over
{(s-1)^2}} ~~~~~~~~~~~~~~~~$$
$$-\sum_{k=1}^\infty {{(s)_{2k}} \over {4^k (2k+1)!}}\left \{
[\psi(s+2k)-\psi(s)]\zeta(s+2k,a)+\zeta'(s+2k,a)\right\}, \eqno(4.19)$$
and apply the definition (0.4).  For Eq. (4.16) we use the property
$${d \over {ds}}(s)_{pk}=(s)_{pk}[\psi(s+pk)-\psi(s)], \eqno(4.20)$$
so that 
$$\left.{d \over {ds}}\right|_{s=1}(s)_{2k}=(2k)!H_{2k}. \eqno(4.21)$$
More generally, we use the expansion
$$\left(a-{1 \over 2}\right)^{1-s}=\sum_{j=0}^\infty {{(-1)^j} \over {j!}} \ln^j \left(
a-{1 \over 2}\right)(s-1)^j,  \eqno(4.22)$$
and apply the product rule to find Eq. (4.18).

\medskip
\centerline{\bf New expression for $\gamma_n(a)$}
\medskip

Based upon Proposition 6, we expect that the corresponding summation expression for $\gamma_k(a)$ will contain a series of $\zeta^{(j)}$ terms to order $k$.  We have
{\newline \bf Proposition 8}.  Let $s(n,m)$ denote the Stirling numbers of the first
kind.  Then we have for Re $a>0$ (i)
$$\gamma_2(a)=-{1 \over 3}\ln^3 (a+1)+{{\ln^2 a} \over a}-\sum_{k=1}^\infty {1 
\over {(k+1)}}\left[(-1)^k \zeta''(k+1,a+1)-{2 \over {k!}}s(k+1,2) \zeta'(k+1,a+1)\right.$$
$$\left. +{2 \over {k!}}s(k+1,3)\zeta(k+1,a+1)\right].  \eqno(5.1)$$
Our expectation is fully borne out, as our proof of part (i) extends to general
$n$ (ii):
$$\gamma_n(a)=-{1 \over {n+1}}\ln^{n+1} (a+1)+{{\ln^n a} \over a}$$
{\small
$$-(-1)^n\sum_{k=1}^\infty {1 \over {(k+1)}}\left[(-1)^k \zeta^{(n)}(k+1,a+1)-{{n!} \over {k!}}\sum_{j=0}^{n-1} {{(-1)^j}
\over {(n-j-1)!}}s(k+1,j+2) \zeta^{(n-j-1)}(k+1,a+1)\right].  \eqno(5.2)$$}

Proof.  We will use a generating function for the Stirling numbers $s(j,k)$,
$${1 \over x}\ln^m (1+x)=m!\sum_{n=m-1}^\infty s(n+1,m) {x^n \over {(n+1)!}}, ~~~~|x|< 1.
\eqno(5.3)$$
Proceeding similarly to the beginning of the proof of Proposition 6, we have generally
$$\gamma_k(a)=-{1 \over {k+1}}\ln^k(a+1)+{{\ln^k a} \over a}+\sum_{j=1}^\infty \int_0^1 \left[{{\ln^k(j+a)} \over {j+a}} - {{\ln^k(x+j+a)} \over {x+j+a}}\right]dx.  \eqno(5.4)$$
Now we apply the expansion (5.3) to write for part (i)
$${{\ln^2 y} \over y}-{{\ln^2 (x+y)} \over {x+y}}=-\sum_{k=1}^\infty {x^k \over y^{k+1}}\left[(-1)^k\ln^2 y+{2 \over {k!}}s(k+1,2)\ln y+{2 \over {k!}}s(k+1,3)\right].  \eqno(5.5)$$
We then substitute this equation into Eq. (5.4) at $k=2$ and perform the integration.
We then perform the summation over $j$, using
$$\sum_{j=1}^\infty {{\ln^p(j+a)} \over {(j+a)^{k+1}}}=(-1)^p \zeta^{(p)}(k+1,a+1),
\eqno(5.6)$$
and part (i) follows.  
Similarly, for part (ii), in place of (5.5) we use
$${{\ln^n y} \over y}-{{\ln^n (x+y)} \over {x+y}}=\sum_{k=1}^\infty {x^k \over y^{k+1}}\left[(-1)^k\ln^n y + {1 \over {k!}}\sum_{j=0}^{n-1} {{n!} \over {(n-j-1)!}}
s(k+1,j+2) \ln^{n-j-1} y \right].  \eqno(5.7)$$
We substitute this equation into Eq. (5.4), perform the integration, perform the summation over $j$ using (5.6), giving Eq. (5.2).  The Proposition is complete.

{\bf Remarks}.  Proposition 6 corresponds to the well known special case of the
harmonic numbers wherein $s(n+1,2)=(-1)^{n+1}n! H_n$.  The connection of Stirling
numbers with sums of generalized harmonic numbers $H_n^{(r)}$ is well known.  For
example, we have $s(n+1,3)=(-1)^n n![H_n^2-H_n^{(2)}]/2$.

As for $\gamma_1(a)$, an integral representation for $\gamma_n(a)$ may be developed
using Eq. (2.8).  We omit such consideration.

If desired, Proposition 8 may be used to write an expression for the differences
$\gamma_n(a)-\gamma_n(b)$.  We have been informed that R. Smith has also obtained Eq.\ (5.2), and studied its rate of convergence \cite{rsmith}.

\medskip
\centerline{\bf Integral expression for $1-\gamma-\gamma_1$, and more}
\medskip

Let $\{x\}=x-[x]$ denote the fractional part of $x$.  It is known that 
$$I_1 \equiv \int_0^1 \left\{ {1 \over x} \right\}dx=1-\gamma,  \eqno(6.1)$$
a result that is reproved and discussed in the Appendix.  We show here
{\newline \bf Proposition 9}.  We have
$$I_2 \equiv \int_0^1 \int_0^1 \left\{{1 \over {xy}}\right\}dxdy = \int_1^\infty
\int_x^\infty {{ \left\{y\right\} } \over {xy^2}}dydx=1-\gamma-\gamma_1.
\eqno(6.2)$$
Although this result is supposedly proved in Ref. \cite{sebah}, we have not
been able to obtain it, and so present our own proofs.  These methods of proof may 
themselves be of independent interest.

Proof 1.  We first establish
{\newline \bf Lemma 2}.  We have
$$\int_x^\infty {{ \left\{y\right\} } \over {y^2}}dy=H_{[x]}-\gamma-\ln x+1
-{{[x]} \over x}.  \eqno(6.3)$$
We have
$$\int_x^\infty {{ \left\{y\right\} } \over {y^2}}dy=\int_{[x]+\left\{x\right\}}^\infty
{{ \left\{y\right\} } \over {y^2}}dy$$
$$=\int_{[x]}^\infty {{ \left\{y\right\} } \over {y^2}}dy
-\int_{[x]}^x{{ \left\{y\right\} } \over {y^2}}dy$$
$$=\sum_{j=[x]}^\infty \int_j^{j+1} {{(y-[y])} \over y^2}dy-\int_{[x]}^x {{(y-[x])}
\over y^2}dy$$
$$=\sum_{j=[x]}^\infty \left[\ln\left({{j+1} \over j}\right)-j\left({1 \over j}-{1 \over
{j+1}}\right)\right]-\ln \left({x \over {[x]}}\right)+[x]\left({1 \over {[x]}}-{1 \over x}\right)
.\eqno(6.4)$$
The sum here is given by (cf. the Appendix)
$$\sum_{j=[x]}^\infty \left[\ln\left({{j+1} \over j}\right)-{1 \over {j+1}}\right]
=\sum_{j=1}^\infty \left[\ln\left({{j+1} \over j}\right)-{1 \over {j+1}}\right]
-\sum_{j=1}^{[x]-1} \left[\ln\left({{j+1} \over j}\right)-{1 \over {j+1}}\right]$$
$$=1-\gamma-\sum_{j=2}^{[x]}\left[\ln j-\ln(j-1)-{1 \over j}\right]$$
$$=1-\gamma-\ln[x]+H_{[x]}-1=H_{[x]}-\ln[x]-\gamma.  \eqno(6.5)$$
Inserting this expression into Eq. (6.4) we obtain the Lemma.

Using the Lemma, we have
$$I_2=\int_1^\infty\left(H_{[x]}-\gamma-\ln x+1-{{[x]} \over x}\right){{dx} \over x}$$
$$=\lim_{M \to \infty}\left\{\int_1^M (1-\gamma-\ln x){{dx} \over x}
+\int_1^M \left(H_{[x]}-{{[x]} \over x}\right){{dx} \over x} \right \}$$
$$=\lim_{M \to \infty} \left[(1-\gamma)\ln M-{1 \over 2}\ln^2 M+\sum_{j=1}^M \int_j^{j+1}
\left(H_{j}-{j \over x}\right) {{dx} \over x}\right]$$
$$=\lim_{M \to \infty} \left\{(1-\gamma)\ln M-{1 \over 2}\ln^2 M+\sum_{j=1}^M \left[H_j
\ln\left({{j+1} \over j}\right)-{1 \over {j+1}}\right]\right\}.  \eqno(6.6)$$
Now the term
$$\sum_{j=1}^M H_j\ln\left({{j+1} \over j}\right)=\sum_{j=1}^M \sum_{k=1}^j {1 \over
k}\ln\left({{j+1} \over j}\right)$$
$$=\sum_{k=1}^M {1 \over k} \sum_{j=k}^M \ln \left({{j+1} \over j}\right)
=\sum_{k=1}^M {1 \over k}[\ln(M+1)-\ln k]$$
$$=H_M \ln(M+1)-\sum_{k=1}^M {{\ln k} \over k}.  \eqno(6.7)$$
Then by Eq. (6.6) we have
$$I_2=\lim_{M \to \infty} \left\{(1-\gamma)\ln M-{1 \over 2}\ln^2 M + H_M \ln(M+1) -\sum_{k=1}^M {{\ln k} \over k} -H_{M+1}+1 \right\}. \eqno(6.8)$$  
We recall the asymptotic form $H_M=\ln M+\gamma+O(1/M)$ as $M \to \infty$ and appeal to
Eq. (0.6) at $a=k=1$:
$$I_2=\lim_{M \to \infty} \left\{\ln M-\gamma\ln M-{1 \over 2}\ln^2 M + \left[\gamma+
\ln M+O\left({1 \over M}\right)\right]\left[\ln M+O\left({1 \over M}\right)\right]\right.$$
$$\left. -\sum_{k=1}^M {{\ln k} \over k} -\left[\gamma+\ln M+O\left({1 \over M}\right) \right]+1 \right\}$$ 
$$=\lim_{M \to \infty} \left\{{1 \over 2}\ln^2 M -\sum_{k=1}^M {{\ln k} \over k}-\gamma+1+O\left({{\ln M} \over M}\right)\right\}$$
$$=1-\gamma-\gamma_1.  \eqno(6.9)$$ 

The alternative form of $I_2$ in Eq. (6.2) results from the change of variable
$(u,v)=(1/xy,x)$, with inverse transformation $(x,y)=(v,1/uv)$, and Jacobian
$$\left|{{\partial(x,y)} \over {\partial(u,v)}}\right|=\left|\begin{array}{cc}
0 & 1\\-{1 \over {u^2 v}} & -{1 \over {u v^2}} \end{array} \right|={1 \over {u^2 v}}.
\eqno(6.10)$$
Then we obtain
$$I_2=\int_0^1\int_{1/v}^\infty {{ \left\{u\right\} } \over {u^2v}}dudv
=\int_1^\infty \int_{t}^\infty {{ \left\{z\right\} } \over {z^2t}}dzdt.
\eqno(6.11)$$
For the second equality we have used the simple transformation
$(v,u)=(1/t,z)$, with corresponding Jacobian
$$\left|{{\partial(x,y)} \over {\partial(u,v)}}\right|=-{1 \over t^2}.  \eqno(6.12)$$

We may obtain a second shorter proof of Proposition 9 by using a known result, namely
the integral representation \cite{ivic}
$$\gamma_k=-\int_1^\infty {1 \over t} \ln^k t ~dP_1(t)
=\int_1^\infty {{\ln^{k-1} t} \over t^2}(k- \ln t) P_1(t)dt -\delta_{k0}/2,
\eqno(6.13)$$
where $\delta_{jk}$ is the Kronecker symbol.  We interchange the order of integration
in the second form of $I_2$ on the right side of Eq. (6.2), giving
$$I_2=\int_1^\infty \int_1^y {{ \left\{y\right\} } \over {xy^2}}dxdy
=\int_1^\infty  {{ \left\{y\right\} } \over {y^2}}\ln y ~dy \eqno(6.14)$$
$$=-\int_0^1 \left\{{1 \over v}\right\} \ln v ~dv.  \eqno(6.15)$$
Therefore, from Eq. (6.14) and Eq. (6.11) at $k=1$ we obtain
$$I_2=\int_1^\infty {{[P_1(w)+1/2]} \over w^2}\ln w ~dw$$
$$=-\gamma-1+\int_1^\infty {{P_1(w)} \over w^2}dw+{1 \over 2}\int_1^\infty {{\ln w} 
\over w^2}dw$$
$$=-\gamma_1-\gamma+{1 \over 2}+{1 \over 2}=-\gamma_1-\gamma+1.  \eqno(6.16)$$

{\bf Remarks}.  Since $0<\{x\} \leq 1$ for $x>0$, we easily have from Eqs. (A.1) and
(6.11) or (6.14) the inequalities $0<I_1 \leq 1$ and $0<I_2 \leq 1$.

We see from Eq. (6.13) that higher order Stieltjes constants may be
obtained from higher dimensional integrals.  For instance, there is a $\gamma_2$
contribution in the integral
$$\int_1^\infty \int_1^z \int_1^z {{ \left\{z\right\} } \over {z^2}}{1 \over {xy}}
dxdydz=\int_1^\infty \int_y^\infty \int_1^z {{ \left\{z\right\} } \over {z^2xy}}
dxdzdy,  \eqno(6.17)$$
and such integrals may be further rewritten with changes of variables.

Indeed, we introduce the integral
$$I_3 \equiv \int_0^1 \int_0^1 \int_0^1 \left\{{1 \over {xyz}}\right\}dxdydz
= \int_1^\infty \int_{1/u}^1 \int_{1/v}^1 {{ \left\{u\right\} } \over {u^2vw}}
dwdvdu, \eqno(6.18)$$
and have 
$$I_3=1-\gamma-\gamma_1-{1 \over 2}\gamma_2.  \eqno(6.19)$$

This is a special case of the following.
{\newline \bf Proposition 10}.  Put
$$I_n \equiv \int_0^1 \int_0^1 \cdots \int_0^1 \left\{{1 \over {x_1x_2\cdots x_n}}\right\}dx_1 dx_2 \cdots dx_n, ~~~~~~ n \geq 1. \eqno(6.20)$$
Then we have
$$I_n=1-\sum_{j=0}^{n-1} {\gamma_j \over {j!}}.  \eqno(6.21)$$

{\bf Corollary 4}.  We have
$$\lim_{n \to \infty} I_n=-\zeta(0)={1 \over 2}.  \eqno(6.22)$$

Proof.  We use the change of variable
$(x_1,x_2,\ldots,x_n)=(u_n,u_{n-1},\ldots,u_2,1/u_1u_2\cdots u_n)$, with Jacobian
$$J=\left|\begin{array}{ccccc}
0 & 0 & \cdots & 0 & 1\\
0 & 0 & \cdots & 1 & 0\\
\vdots & 0 & 1 & 0 & \vdots\\
0 & 1 & \cdots & 0 & 0\\
-{1 \over {u_1^2 u_2\cdots u_n}} & 0 & \cdots & 0 & 0 
\end{array} \right|={\varepsilon_n \over {u_1^2 u_2 \cdots u_n}},  \eqno(6.23)$$
where $\varepsilon_n =\pm 1$.  Specifically, if $n$ is of the form $4m$ or $4m+1$,
$\varepsilon_n=-1$, and $\varepsilon_n=1$ otherwise.  
Then
$$I_n =\varepsilon_n \int_1^\infty \int_{1/u_2}^1 \cdots \int_{1/u_{n-1}}^1 {{ \{u_1\} } \over u_1^2} {{ \prod_{i=n}^1 du_i} \over {\prod_{i=2}^n u_i}}.  \eqno(6.24)$$
We then multiply integrate to find
$$I_n={1 \over {(n-1)!}}\int_1^\infty {{ \{u_1\} } \over u_1^2} \ln^{n-1} u_1 ~du_1.
\eqno(6.25)$$
We now note from Eq. (6.13)
$$\sum_{k=0}^{n-1} {\gamma_k \over {k!}}=-{1 \over {(n-1)!}}\int_1^\infty \Gamma(n,
\ln t) ~dP_1(t)$$
$$=-{1 \over {(n-1)!}}\int_1^\infty {{\ln^{n-1} t} \over t^2}P_1(t) ~dt+{1 \over 2},
\eqno(6.26)$$
where we applied \cite{grad} (p. 941), integrated by parts, and used $\Gamma(n,0)=
(n-1)!$ for $n \geq 1$ an integer.  We then have the Proposition, as
$$\sum_{k=0}^{n-1} {\gamma_k \over {k!}}=-{1 \over {(n-1)!}}\int_1^\infty 
{{\ln^{n-1} t} \over t^2}\left(\{t\}-{1 \over 2}\right)dt+{1 \over 2}$$
$$=-{1 \over {(n-1)!}}\int_1^\infty \{t\} {{\ln^{n-1} t} \over t^2}dt+1.  \eqno(6.27)$$

{\bf Remarks}.
We have found that the subjects of Propositions 9 and 10 have recently been of interest elsewhere \cite{furdui}.  It would be interesting to have a probabilistic argument for
Corollary 4.



\medskip
\centerline{\bf Summary}
\medskip

We have obtained new explicit analytic results for the Stieltjes coefficients
including series representations and summatory relations.   Other integral 
representations based upon the properties of Dirichlet $L$-functions provide the
difference of Stieltjes coefficients at rational arguments, and these give inequalities.  
Our results have implications for other coefficients 
of analytic number theory and other fundamental mathematical constants.  

\bigskip
\centerline{\bf Acknowledgement}
\medskip
I thank R. Kreminski for access to high precision values of the Stieltjes coefficients.
I thank R. Smith for correspondence and various numerical verifications.



\pagebreak
\bigskip
\centerline{\bf Appendix:  Integral expression for $1-\gamma$}
\medskip

We here show that
{\newline \bf Proposition A1}.  We have
$$I_1 \equiv \int_0^1 \left\{ {1 \over x} \right\}dx=\int_1^\infty {{ \left\{x\right\} } \over {x^2}}dx =1-\gamma.  \eqno(A.1)$$

Proof.  We recall that $\{x\}=P_1(x)+1/2=x-[x]$, giving
$$I_1=\int_1^\infty {{\left\{x\right\} } \over {x^2}}dx=\sum_{j=1}^\infty \int_j^{j+1}
\left({1 \over x}-{j \over x^2}\right)dx$$
$$=\sum_{j=1}^\infty \left[\ln\left({{j+1} \over j}\right)-{1 \over {j+1}}\right]$$
$$=\lim_{N\to \infty} \sum_{j=1}^N \left[\ln(j+1)-\ln j -{1 \over {j+1}}\right]$$
$$=\lim_{N \to \infty}[\ln N-H_N+1]=1-\gamma, \eqno(A.2)$$
where we used the telescoping nature of the sum in the next-to-last line.

As a variation on this proof, we may note that a sum above is a case of the 
summation \cite{hansentable} (44.9.1, p. 290)
$$\sum_{k=1}^\infty \left[\ln\left({{k+x} \over k}\right)-{x \over k}\right]
=-\gamma x +\ln \Gamma (x+1).  \eqno(A.3)$$
We then obtain
$$\sum_{j=1}^\infty \left[\ln\left({{j+1} \over j}\right)-{1 \over {j+1}}\right]$$
$$=\sum_{j=1}^\infty \left[\ln\left({{j+1} \over j}\right)-{1 \over j}+{1 \over j}
-{1 \over {j+1}}\right]$$
$$=-\gamma + \ln \Gamma(2)+\sum_{j=1}^\infty\left({1 \over j}-{1 \over {j+1}}\right)
=1-\gamma.  \eqno(A.4)$$
Similarly, this sum may be found at $x=0$ in \cite{hansentable} (44.9.4, p. 290) or
at $x=1$ in \cite{hansentable} (44.9.5, p. 290).

\pagebreak

\end{document}